\documentclass[referee]{raa}            

\usepackage{graphicx,times}             
\usepackage{natbib}
\usepackage{amssymb,amsmath}
\usepackage{color}
\bibpunct{(}{)}{;}{a}{}{,}

\usepackage[a4paper=true,dvipdfm=true,pagebackref=true]{hyperref}
\hypersetup{colorlinks = true, linkcolor = green, anchorcolor = red, citecolor = blue, filecolor = red, pagecolor = red, urlcolor = red}

\begin{document}

   \title{Polarization of GRB Prompt Emission and its Application to POLAR's Data
}

   \volnopage{Vol.0 (20xx) No.0, 000--000}      
   \setcounter{page}{1}          

   \author{Mi-Xiang Lan
      \inst{1}
   \and Xue-Feng Wu
      \inst{2}
   \and Zi-Gao Dai
      \inst{3,4}
   }

   \institute{Center for Theoretical Physics and College of Physics, Jilin University, Changchun, 130012, China; {\it  lanmixiang@jlu.edu.cn}\\
        \and
             Purple Mountain Observatory, Chinese Academy of Sciences, Nanjing 210008, China;  {\it  xfwu@pmo.ac.cn}\\
        \and
             School of Astronomy and Space Science, Nanjing University, Nanjing 210093, China;  {\it  dzg@nju.edu.cn}\\
        \and
             Key Laboratory of Modern Astronomy and Astrophysics (Nanjing University), Ministry of Education, China \\
\vs\no
   {\small Received~~20xx month day; accepted~~20xx~~month day}}

\abstract{Synchrotron emission polarization is very sensitive to the magnetic field configuration. Recently, polarization of synchrotron emission with a mixed (SM) magnetic field in Gamma-ray burst (GRB) afterglow phase had been developed. Here, we apply these SM models to GRB prompt phase and compare their polarization properties with that of synchrotron emission in purely ordered (SO) magnetic field. We find that the polarization properties in a SM model are very similar to these in a corresponding SO model (e.g., synchrotron emission in a mixed magnetic field with an aligned ordered part (SMA) and synchrotron emission with a purely ordered aligned magnetic field (SOA)), only with a lower polarization degree (PD). We also discuss the statistical properties of the models. We find PDs of the simulated bursts are concentrated around $25\%$ for both SOA and synchrotron emission in a purely ordered toroidal magnetic field (SOT), while they can range from $0\%$ to $25\%$ for SMA and synchrotron emission in a mixed magnetic field with a toroidal ordered part (SMT), depending on $\xi_B$ value, i.e., the ratio of magnetic reduction of the ordered magnetic field over that of random magnetic field. From statistics, if PDs of majority GRBs are non-zero, then it favours SO and SM models. Further, if there are some bright GRBs with a prominently lower PDs than that of the majority GRBs, it favours SOT (SMT) models; if all the bright GRBs have comparable PDs with the majority ones, it favours SOA (SMA) models. Finally, we apply our results to POLAR's data and find that $\sim10\%$ time-integrated PDs of the observed bursts favor SMA and SMT models, and $\xi_B$ parameter of these bursts is constrained to be around 1.135.
\keywords{gamma-ray burst: general --- magnetic fields --- polarization --- radiation mechanisms: nonthermal}
}

   \authorrunning{M. X. Lan, X. F. Wu \& Z. G. Dai }            
   \titlerunning{Polarization of GRB Prompt Emission and its Application to POLAR's Data }  

   \maketitle

%
%
\section{Introduction}           
\label{sect:intro}

Gamma-ray bursts (GRBs) are luminous $\gamma$-ray transients at cosmological distances. The non-thermal spectra of most of the observed GRBs are described by the Band function (Band et al. 1993), of which two power laws are jointed at a break energy $E_{p,obs}$ in the $\nu f_{\nu}$ spectrum. Three popular models of GRB prompt phase had been proposed so far, i.e., the internal shock model (Rees \& M\'esz\'aros 1994; Narayan, Paczynski \& Piran 1992), magnetic reconnection model (Giannios 2008; Zhang \& Yan 2011; Beniamini \& Granot 2016; Granot 2016) and photospheric model (Thompson 1994; Eichler \& Levinson 2000; M\'esz\'aros \& Rees 2000; Rees \& M\'esz\'aros 2005; Lazzati et al. 2009; Beloborodov 2011; Pe'er \& Ryde 2011; Mizuta et al. 2011; Nagakura et al. 2011; Ruffini et al. 2013; Xu et al. 2012; B\'egu\'e et al. 2013; Lundman et al. 2013; Lazzati et al. 2013). The predicted features of all these three models can match the observations (Uhm \& Zhang 2014; Zhang \& Zhang 2014; Pe'er, M\'esz\'aros \& Rees 2005, 2006; Rees \& M\'esz\'aros 2005; Abramowicz, Novikov \& Paczynski 1991; Pe'er 2008). Even for two-decade studies, the emission mechanism and magnetic field configuration (MFC) during GRB prompt phase have remained mysterious. Polarization strongly depends on these two factors and can conversely be used as a probe (Granot 2003; Toma et al. 2009; Lan, Wu \& Dai 2016a; Lan et al. 2019).

In fact, there are several $\gamma$-ray polarimeters in commission (Winkler et al. 2003; Hitomi Collaboration et al. 2018) and a number of prompt polarization data have been accumulated. Most of the observed GRBs have a lower limit of polarization degree (PD) and the minimum lower limit is about $30\%$ (Willis et al. 2005; McGlynn et al. 2007; G$\ddot{o}$tz et al. 2013, 2014). The observed PD values of GRBs 100826A, 110301A and 110721A are $27\pm11\%$, $70\pm22\%$ and $84^{+16}_{-28}\%$, respectively (Yonetoku et al. 2011, 2012). These PD observations mentioned above suggest that GRB prompt emissions are highly polarized.

Recently, the POLAR team published their polarization observation results of five GRBs (Zhang et al. 2019). Different from but consistent with the lower limit of the previous results, POLAR's data show that most of the bright GRBs may be moderately polarized, with a PD of $\sim10\%$. Another observational quantity of polarization is its direction, usually depicted by polarization angle (PA). Up till now, the measurements of PA are very rare (McGlynn et al. 2007; Yonetoku et al. 2011, 2012; Burgess et al.2019). McGlynn et al. (2007) analyzed the data of GRB 041219A and found PAs for both the 12 s and 66 s time intervals are constant. In GRBs 110301A and 110721A, PAs also keep roughly as constant, while in GRB 100826A PAs for the two bright intervals have a roughly $90^\circ$ difference (Yonetoku et al. 2011, 2012). Recently, Burgess et al.(2019) reanalyzed the POLAR's observational data of GRB 170114A and found a gradually evolving PA for this burst.

Polarizations in GRBs have been widely studied, including its properties with different emission mechanism (Shaviv \& Dar 1995; Sari 1999; Gruzinov 1999), with different MFCs (Sari 1999; Granot \& K\"{o}nigl 2003; Toma et al. 2009; Lan et al. 2018, 2019) and with various jet structure (Rossi et al. 2004; Wu et al. 2005). The emission mechanism for GRB prompt phase can be synchrotron or inverse Compton (Wang et al. 2019; Fraija et al. 2019). Literally, polarizations with three kinds of MFCs have been studied (Sari 1999; Granot \& K\"{o}nigl 2003; Toma et al. 2009). An ordered aligned MFC usually originates from a perpendicular rotator of a magnetar (Spruit et al. 2001) and an ordered toroidal MFC might be generated through the Blandford-Znajek mechanism of a black hole (Spruit et al. 2001), while a three-dimensional (3D) anisotropic random magnetic field might be generated by a shock or by magnetic reconnection. Recently, polarizations of synchrotron emission with a 3D mixed (SM) MFC at GRB afterglow phase had been discussed by Lan et al. (2019) and Stokes parameters in a total magnetic field, including both the ordered and random components, were considered.

Because the polarization properties of synchrotron emission are very sensitive to the MFC, polarizations of GRB prompt phase with the newly developed SM models are investigated in this paper. We discuss the polarization properties of synchrotron emission with these new MFCs in GRB prompt phase and compare their results with these of the traditional MFCs (i.e., purely ordered MFC and 2D random MFC confined in the shock plane). This paper is arranged as follows. In Section 2, we propose our polarization models. Numerical results of these models are exhibited in Section 3. In Section 4, we calculate the statistical properties of GRB polarization. Then we apply our models to POLAR's data in Section 5. Conclusions and discussion are presented in Section 6.


\section{Polarization Models}
\label{sect:Obs}

We consider the emission of an ultra-relativistic jet, located at redshift $z$. The emission region of the jet is assumed to be a thin shell and it is optically thin to $\gamma$-rays. For an observer with viewing angle $\theta_V$, the spectral fluence of the jet can be expressed as follows (Toma et al. 2009; Ioka \& Nakamura 2001; Granot et al. 1999; Woods \& Loeb 1999).
\begin{equation}
F_{\nu}=\frac{1+z}{d_L^2}R^2\int_0^{\theta_j+\theta_V}\sin\theta d\theta \mathcal{D}^2f(\nu') \int_{-\Delta\phi}^{\Delta\phi}d\phi A_0,
\end{equation}
$d_L$ is the luminosity distance of the source, $R$ is the radius of the emission region. $\theta_j$ is jet half-opening angle, $\mathcal{D}=1/\Gamma(1-\beta\cos\theta)$ is the Doppler factor, $\Gamma$ and $\beta$ are the bulk Lorentz factor and the velocity of the jet in units of speed of light, $\theta$ is the angle between the line of sight and local radial direction, $\nu'=\nu_{obs}(1+z)/\mathcal{D}$ and $\nu_{obs}$ is the observational frequency. $\phi$ is the angle in the plane of sky between the projection of jet axis and the projection of the local fluid velocity direction. The expression of $\Delta\phi$ can be found in Toma et al. (2009) and Lan et al. (2016a). $A_0$ is a normalization factor, with units of erg/cm$^2/$Hz/str. The primed and unprimed quantities are in the comoving frame and the observer frame, respectively. The spectrum ($f(\nu')=g(x)$) of GRB prompt emission is assumed to be described by Band function (Band et al. 1993).
\begin{equation}
g(x)=\begin{cases}
x^{-\alpha_s}e^{-x}, & \text{$x<\beta_s-\alpha_s$}, \\ x^{-\beta_s}(\beta_s-\alpha_s)^{\beta_s-\alpha_s}e^{\alpha_s-\beta_s}, & \text{$x\geq\beta_s-\alpha_s$},
\end{cases}
\end{equation}
where $x=\nu'/\nu'_0$, $\nu'_0$ is the comoving break energy of the Band spectrum, $\alpha_s$ and $\beta_s$ are the low-energy and high-energy spectral indices. Then the spectral index $\tilde{\alpha}$ can be expressed as
\begin{equation}
\tilde{\alpha}=\begin{cases}
\alpha_s, & \text{$x<\beta_s-\alpha_s$}, \\ \beta_s, & \text{$x\geq\beta_s-\alpha_s$},
\end{cases}
\end{equation}

Here, we only consider the linear polarization. The Stokes parameters, which describe the linear polarization, can be expressed as
\begin{equation}
Q_{\nu}=\frac{1+z}{d_L^2}R^2\int_0^{\theta_j+\theta_V}\sin\theta d\theta\mathcal{D}^2f(\nu')\int_{-\Delta\phi}^{\Delta\phi}d\phi A_0\Pi_p\cos(2\chi_p),
\end{equation}
\begin{equation}
U_{\nu}=\frac{1+z}{d_L^2}R^2\int_0^{\theta_j+\theta_V}\sin\theta d\theta\mathcal{D}^2f(\nu')\int_{-\Delta\phi}^{\Delta\phi}d\phi A_0\Pi_p\sin(2\chi_p),
\end{equation}
$\Pi_p$ and $\chi_p$ are the local PD and PA, respectively.

Then if both Stokes parameters $Q_\nu$ and $U_\nu$ are non-zero, PD ($\Pi$) and PA ($\chi$) of the jet emission are expressed as
\begin{equation}
\Pi=\frac{\sqrt{Q_\nu^2+U_\nu^2}}{F_\nu}
\end{equation}
\begin{equation}
\chi=\frac{1}{2}\arctan\left(\frac{U_\nu}{Q_\nu}\right)
\end{equation}
If one of the Stokes parameters $U_\nu$ is zero, the PD of the jet emission is defined as follows (Sari 1999).
\begin{equation}
\Pi=\frac{Q_\nu}{F_\nu}
\end{equation}
Here, the absolute value of $\Pi$ represent the magnitude of the polarization. Its sign indicates the polarization direction, polarization direction with $\Pi>0$ will have a $90^\circ$ difference with that of $\Pi<0$.

Here, we consider three classes of MFCs, large-scale ordered (Granot \& K\"{o}nigl 2003; Graont 2003; Toma et al. 2009), mixed (Lan et al. 2019) and random (Sari 1999; Gruzinov 1999; Toma et al. 2009). There are three kinds of ordered magnetic field discussed in the literature, i.e., aligned (Granot \& K\"{o}nigl 2003), toroidal (Toma et al. 2009) and radial (Graont 2003), and the mixed magnetic field consists of ordered and random part, we therefore discuss three subclasses of mixed MFCs with different ordered parts, i.e., aligned + random, toroidal + random and radial + random (Lan et al. 2019). Same as that in Lan et al. (2019), the random part in the mixed MFC is assumed to be isotropic in 3 dimensional space. We also consider two kinds of random MFCs, i.e., 2 dimensional random magnetic field confined in the shock plane (Toma et al. 2009) and 3 dimensional random magnetic field isotropic in space (Lan et al. 2019). Then we have 4 new polarization models, the synchrotron emission in three mixed MFCs with different ordered component (denoted as SMA, SMT and SMR) and synchrotron emission in a 3 dimensional random magnetic field (SR3). We will compare polarization predictions in GRB prompt phase of these 4 new models with that of synchrotron emission in three ordered magnetic fields (denoted as SOA, SOT and SOR) and synchrotron emission in a 2 dimensional random magnetic field (SR2). The corresponding formulas are shown in the Appendix.

\section{Numerical Results for Different Polarization Models}
\label{sect:data}
We take a set of fixed parameters: $\Gamma=100$, $\alpha_s=-0.2$, $\beta_s=1.2$, $E_{p,obs}=350$ keV, $\xi_B=1.5$, $\delta_a=\pi/6$, $z=1$, $\theta_j=0.1$ rad, $\theta_V=0.5\theta_j$ and $h\nu_{obs}=250$ keV. $E_{p,obs}$ is the break energy of the Band spectrum in the observer frame, $\delta_a$ is the orientation of the aligned magnetic field. Without special illustration, parameters used in the following calculations will take the above fixed value. Since there are several parameters that affect the polarization properties, we have discussed their effects in the following. It is interesting to compare the polarization properties in a mixed magnetic field with that in a corresponding ordered magnetic field (e.g., SMA and SOA), we therefore show the results of SM and SO models together.

The statistical properties will be discussed below, the observational angle $\theta_V$ for each GRB would be different, therefore it is necessary to discuss the polarization properties evolving with $\theta_V$, of which are shown in Figs. 1, 2 and 7. Fig. 1 presents the polarization evolution for 6 models, including SMA, SOA, SMT, SOT, SMR and SOR. Here we define $q_{obs}\equiv\theta_V/\theta_j$. The profile of the PD curve for a SM model ($\xi_B=1.5$) is very similar to that with a corresponding SO model (i.e., SMA and SOA, SMT and SOT). The difference is that PD values of SM models are lower than that of SO models as expected.

   \begin{figure}
   \centering
   \includegraphics[width=\textwidth, angle=0]{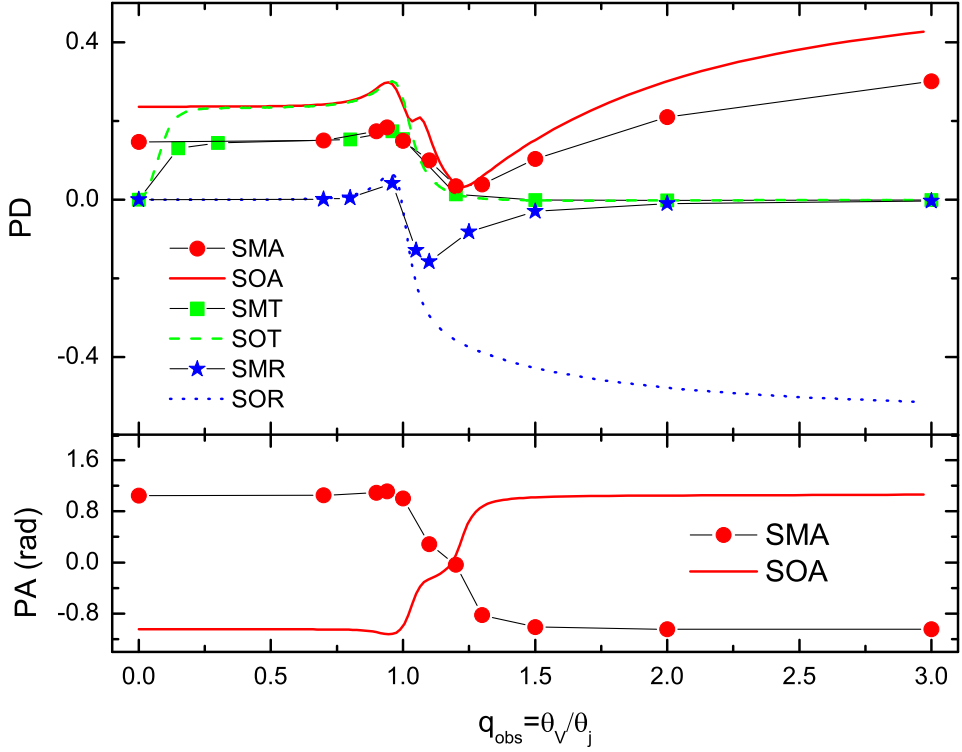}
   \caption{Polarization evolutions with $q_{obs}$ for different SO and SM models. The upper panel shows the PD evolutions and the lower panel corresponds to the PA evolutions. The red solid line is for the SOA model, The green dashed line corresponds to SOT model, the blue dotted line is responsible for the SOR model. The red circles, green diamonds, blue stars are our calculation points and correspond to the SMA, SMT and SMR models, respectively. }
   \label{Fig1}
   \end{figure}

For SOT model, the profile of our PD curve is very similar to that of Toma et al. (2009), but our PD values are lower and the reason for this will be discussed in the following Conclusions and Discussion section. PDs for SOT and SMT models are both 0 when $q_{obs}=0$ because of axial symmetry. Then PDs of these two models rise quickly with $q_{obs}$, because jet axis will move from the center of the observational cone (i.e., $1/\Gamma$ cone) to the edge, leading to more and more incomplete magnetic lines in $1/\Gamma$ cone and hence to an increasing PD. When $1/\Gamma\theta_j<q_{obs}<1-1/\Gamma\theta_j$, PD curves of both SOT and SMT models reach a plateau because the MFC for SOT model or the ordered part of MFC for SMT model in $1/\Gamma$ cone is approximately aligned, this is also the reason for that PD plateau of SOT (SMT) model coincides with that of SOA (SMA) model. For both SOA and SMA models, even when $q_{obs}=0$, their PDs are non-zero because the MFC for SOA model or the ordered part of magnetic field for SMA model in $1/\Gamma$ cone is aligned when the jet axis coincides with the line of sight.

Since the asymmetry due to geometry will increase when $1/\Gamma$ cone crosses the jet edge, PDs will reach a small peak when $q_{obs}$ is slightly smaller than 1 for SMA, SOA, SMT and SOT models. Then PDs begin to decrease for these four models. Beyond some $q_{obs}$, which is slightly larger than $1+1/\Gamma\theta_j$, PDs begin to rise for SMA and SOA models, while they continue to decrease for SMT and SOT models. The evolution trends of PDs for SMA and SOA models behave very different from these for SMT and SOT models. In general, the increasing PDs will indicate an increasing asymmetry of the system and vice versa, here it is some what hard for us to test the asymmetries of these systems. PAs of both SOA and SMA models evolve gradually when $1/\Gamma$ cone crosses the jet cone (roughly when $1-1/\Gamma\theta_j<q_{obs}<1.5$) and keep as constant for other $q_{obs}$.

We notice that with the increase of $q_{obs}$, the profiles of the PD curves for SOR and SMR ($\xi_B=1.5$) models are not similar in our Fig. 1. With the increase of $q_{obs}$, PD curve for SOR model converges to $PD\sim50\%$, while it approaches 0 for SMR model ($\xi_B=1.5$). To examine our results for SOR and SMR models, we then calculate the $q_{obs}-PD$ curves for SMR model with different $\xi_B$ values, which is shown in Fig. 2. When $\xi_B=0$, the magnetic field is 3D isotropic in space, which will lead to no net polarization as shown in Fig. 2. When $\xi_B=30\gg1$, the magnetic field is dominated by the radial component and the PD curve approaches that of SOR model (i.e., $\xi_B\rightarrow\infty$). Therefore, our results for SR3 ($\xi_B=0$), SMR ($0<\xi_B<\infty$) and SOR ($\xi_B\rightarrow\infty$) models are consistent. In Granot (2003), polarization properties of SOR model had been discussed, which is shown in Fig. 4 of his paper. The profile of our PD curve for SOR model is very similar to that of $y_j=(\Gamma\theta_j)^2=100$ shown in Fig. 4 of Granot (2003). Because our coordinate system has rotated by $90^\circ$ relative to that in Granot (2003), the sign of our PDs is opposite compared with that of Granot (2003) at same $q_{obs}$. For SMR and SOR models, large PD will be obtained for off-axis observations.

   \begin{figure}
   \centering
   \includegraphics[width=\textwidth, angle=0]{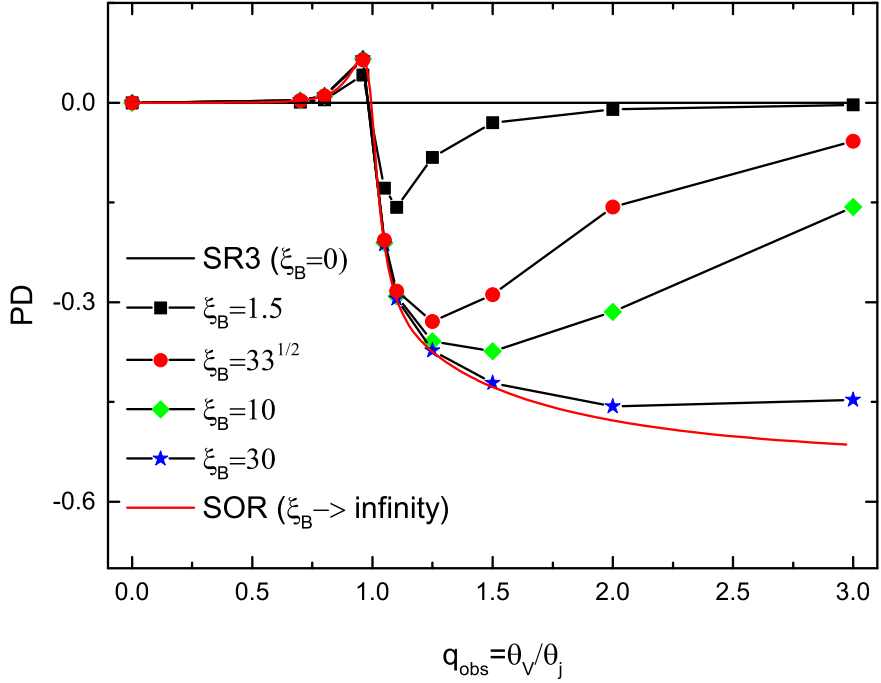}
   \caption{PD evolution with $q_{obs}$ for the SMR model. The black solid line is for the SR3 model and the red solid line corresponds to SOR model, of which are the limit of the SMR model with $\xi_B=0$ and $\xi_B\rightarrow\infty$, respectively. }
   \label{Fig1}
   \end{figure}

The polarization curves evolving with $\theta_j$ for SO and SM models are shown in Fig. 3. The profiles of PD curves are very similar for a SM model and a corresponding SO model. Only the PD values of the SM model are lower compared with that of the corresponding SO model as expected. PDs initially decay with $\theta_j$ and then keep roughly as constant when $\theta_j>0.1$ rad for both SOA and SMA models, While they increase to a peak, then decay slightly and finally keep roughly as constant for SOT, SMT, SOR and SMR models. For SOA and SMA models, the decaying PD is due to loss of observational geometric asymmetry with the increasing jet half-opening angle $\theta_j$. When $\theta_j\gg1/\Gamma$ (e. g., $\theta_j\sim0.1$ rad $\gg1/\Gamma=0.01$), the asymmetry of the system is dominated by the asymmetry of the aligned magnetic field and the asymmetry due to the observational geometry is negligible, so PD values of SOA and SMA models keep roughly as constant when $\theta_j>0.1$ rad. For SOT and SMT models, although the observational geometric asymmetry is decreasing, the magnetic field in $1/\Gamma$ cone becomes more ordered with the increase of $\theta_j$, which lead to an increasing PD initially. Then when $\theta_j\gg1/\Gamma$, the ordered part of the magnetic field in $1/\Gamma$ cone (for SMT) is approximately aligned and the asymmetry of magnetic field in $1/\Gamma$ cone reaches its maximum value and keeps roughly unchanged, leading to a roughly constant PD when $\theta_j>0.1$ rad. Because a toroidal magnetic field in $1/\Gamma$ cone will approach an aligned case when $\theta_j\gg1/\Gamma$, PD of SOT (SMT) model will approach that of SOA (SMA) model at large $\theta_j$ values. PAs of both SOA and SMA models keep as constant with $\theta_j$.

   \begin{figure}
   \centering
   \includegraphics[width=\textwidth, angle=0]{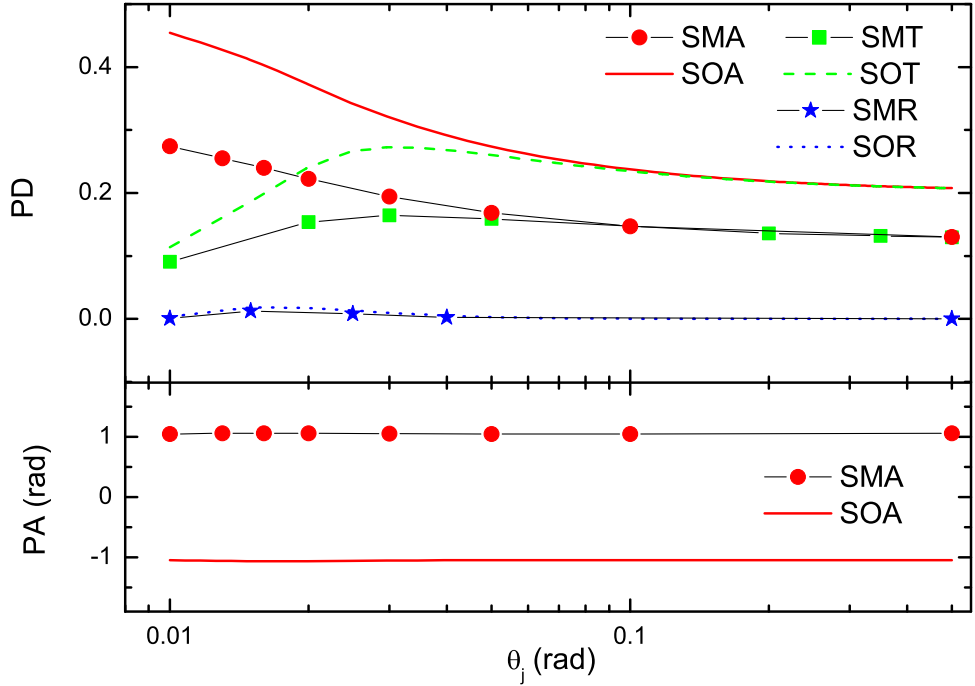}
   \caption{Same as Fig. 1, but for polarization evolutions with jet half-opening angle $\theta_j$. }
   \label{Fig1}
   \end{figure}

Polarization properties of SO and SM models evolving with $\Gamma$ are illustrated in Fig. 4. PDs for SOR and SMR models are roughly 0 for the observational geometry of $\theta_j=0.1$ rad and $q_{obs}=0.5$, no matter which value the bulk Lorentz factor $\Gamma$ takes. The profiles of the PD curves are very similar for SMA, SOA, SMT and SOT models, i.e., PD decays slightly when $\Gamma<100$ due to loss of observational geometric asymmetry and then keeps roughly as a constant after $\Gamma=100$ due to the dominated asymmetry of magnetic field in $1/\Gamma$ cone. PD values for SMA and SMT models are lower than that of SOA and SOT models as expected. For small $\Gamma$ value (e.g. $\Gamma=50$), PDs of SOT (SMT) are slightly lower than that of SOA (SMA), because the MFC in $1/\Gamma$ cone is slightly less ordered for a toroidal magnetic field case than that for an aligned case when $1/\Gamma$ cone is relatively large. Then with an increase of $\Gamma$, the MFC in $1/\Gamma$ cone of a toroidal case approaches that of an aligned case and PD curve of SOT (SMT) model will converge to that of SOA (SMA) when $\Gamma>100$. PAs for both SMA and SOA models keep as constant with $\Gamma$.

   \begin{figure}
   \centering
   \includegraphics[width=\textwidth, angle=0]{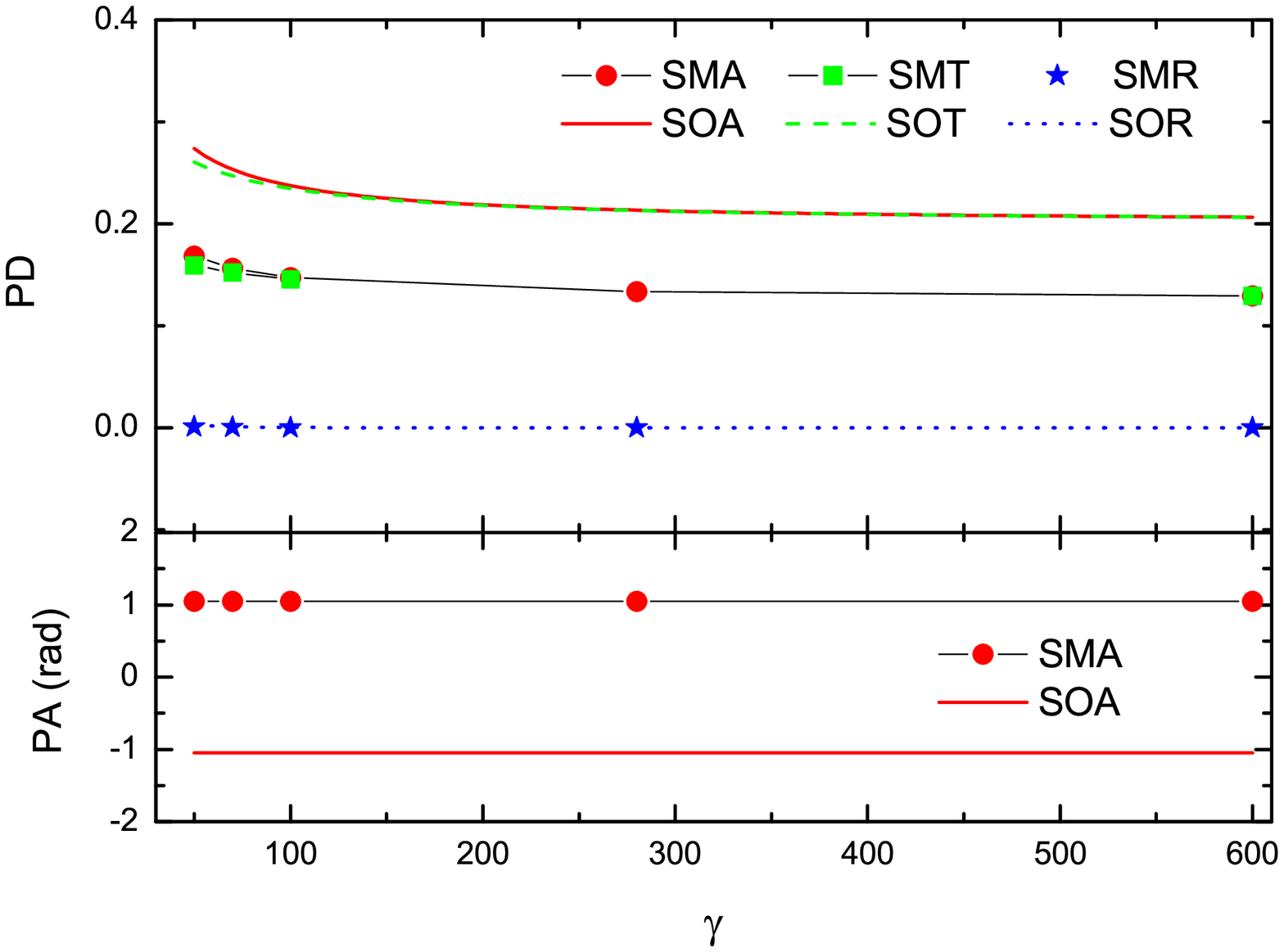}
   \caption{Same as Fig. 1, but for polarization evolutions with jet bulk Lorentz factor $\Gamma$. }
   \label{Fig1}
   \end{figure}

Fig. 5 shows the polarization evolutions with $\nu_{obs}$ for SO and SM models. For SOR and SMR models, PDs are roughly 0 for different $\nu_{obs}$ values with $\theta_j=0.1$ rad and $q_{obs}=0.5$. The profiles of PD curves are very similar for SOA, SMA, SOT and SMT models. PDs are roughly constant for these four models when $h\nu_{obs}<100$ keV, then they increase slightly when $h\nu_{obs}>100$ keV. Because with increasing of $\nu_{obs}$, $x$ value also increases. When $x\geq\beta_s-\alpha_s$, the spectral index $\tilde{\alpha}$ will switch from $\alpha_s$ to $\beta_s$. Since with the increasing of the spectral index $\tilde{\alpha}$, local PD $\Pi_0$ also increases. Therefore, PDs will increase after some observational frequency. PDs for SMA and SMT models are usually smaller than that of SOA and SOT models as expected. PAs for both SMA and SOA models keep as constant with $\nu_{obs}$.

   \begin{figure}
   \centering
   \includegraphics[width=\textwidth, angle=0]{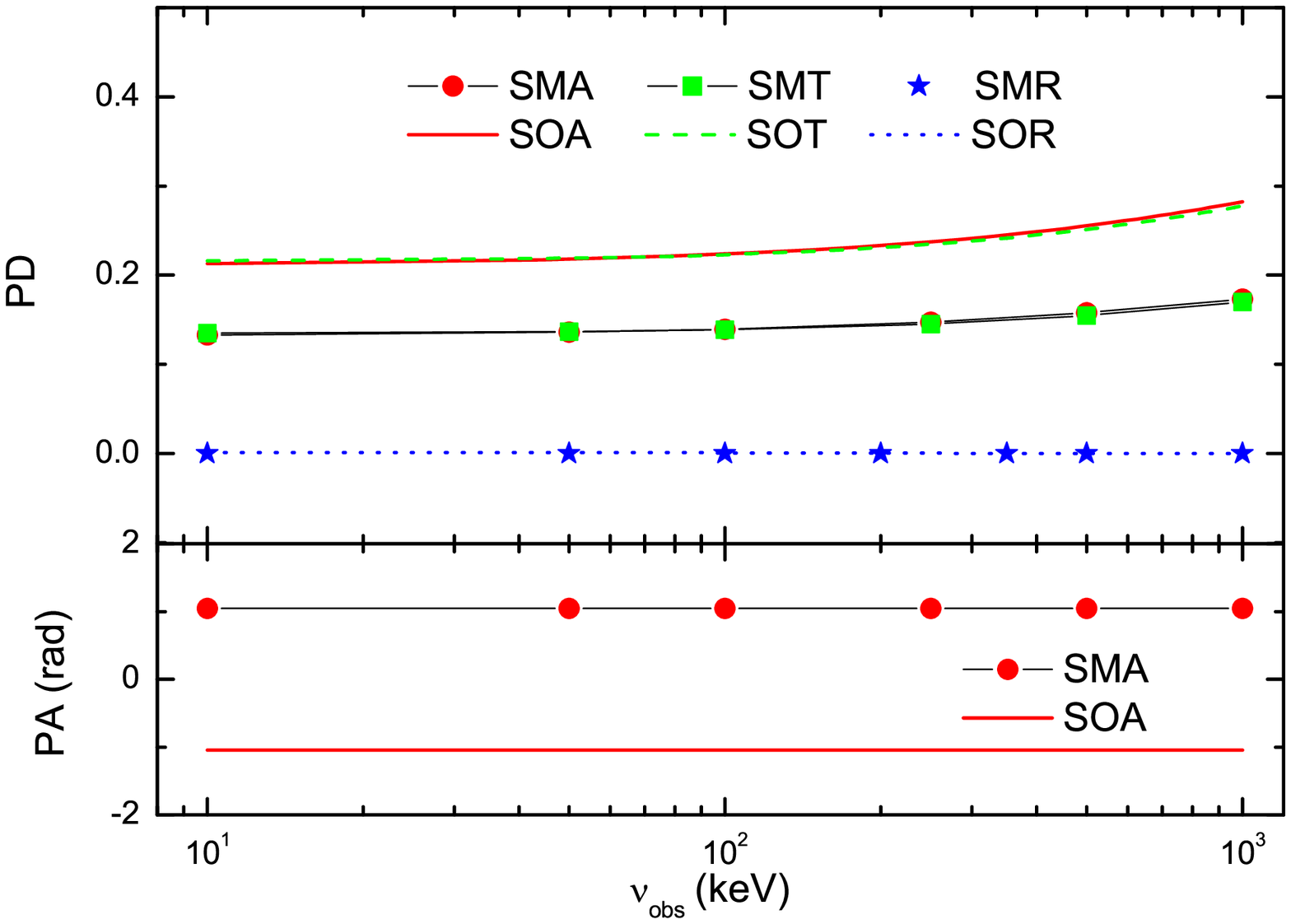}
   \caption{Same as Fig. 1, but for polarization evolutions with observational frequency $\nu_{obs}$. }
   \label{Fig1}
   \end{figure}

Polarization evolutions with peak energy $E_{p,obs}$ for SO and SM models are exhibited in Fig. 6. And also PDs of both SOR and SMR models are roughly 0 for different peak energy. For SOA, SMA, SOT and SMT models, the profiles of their PD curves are similar, i.e., PDs are initially decreasing with $E_{p,obs}$ and then keep roughly as constant after $E_{p,obs}>h\nu_{obs}=250$ keV. Because with the increase of $E_{p,obs}$, for the same observational frequency, $x$ will decrease. When $x<\beta_s-\alpha_s$, the spectral index $\tilde{\alpha}$ will switch from $\beta_s$ to $\alpha_s$. With a decreasing of $\tilde{\alpha}$, local PD $\Pi_0$ also decreases and then leads to a decreasing PD with $E_{p,obs}$ at the beginning. With the increase of $E_{p,obs}$, especially when $x<\beta_s-\alpha_s$ stands for all the fluid elements (approximately at $E_{p,obs}>h\nu_{obs}=250$ keV) , PD of the jet emission will keep as a constant. PAs of SOA and SMA models are both constant with $E_{p,obs}$.

   \begin{figure}
   \centering
   \includegraphics[width=\textwidth, angle=0]{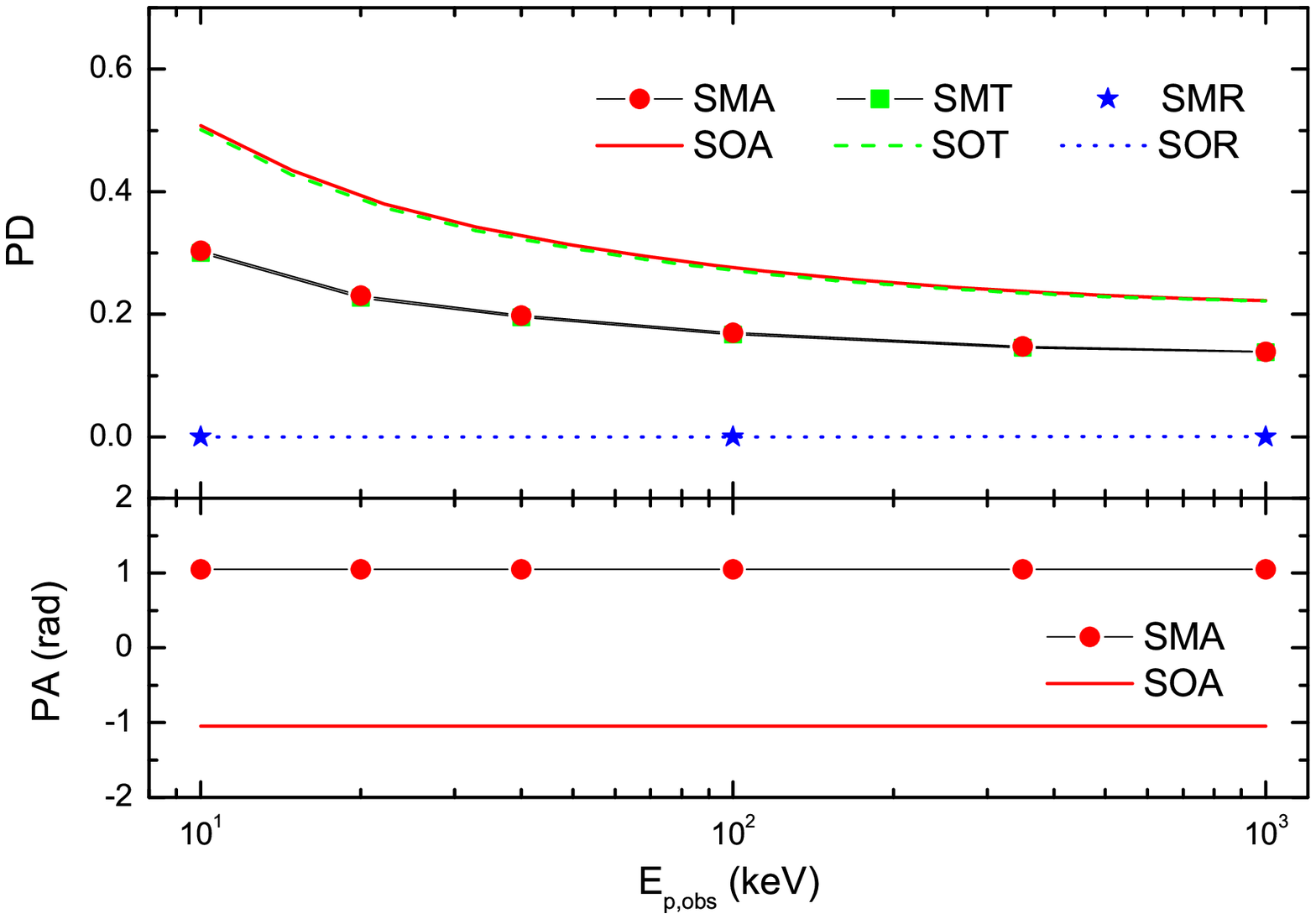}
   \caption{Same as Fig. 1, but for polarization evolutions with peak energy $E_{p,obs}$. }
   \label{Fig1}
   \end{figure}

The $q_{obs}-PD$, $\theta_j-PD$, $\Gamma-PD$, $\nu_{obs}-PD$ and $E_{p,obs}-PD$ curves for SR and CD models are shown in Fig. 7. PD for SR3 model is always 0 independent of various parameters. There are two PD peaks for both SR2 and CD models in $q_{obs}-PD$ figure. These peaks are reached around $q_{obs}=1$, with one peak located roughly at $q_{obs}=\sqrt{1-1/2y_j}-1/\sqrt{2y_j}<1$ and the other at $q_{obs}=\sqrt{1-1/2y_j}+1/\sqrt{2y_j}>1$ with $y_j\equiv(\Gamma\theta_j)^2$ and $1/\Gamma\ll\theta_j$. We also notice that PDs for SR2 and CD models have the opposite signs (if they are both non-zero) at same $q_{obs}$. The absolute value of PD for CD model is higher than that for SR2 model. PAs of both SR2 and CD models change abruptly by $90^\circ$ approximately when $q_{obs}\sim1$ for $1/\Gamma\ll\theta_j$ case. PD values for these three models are roughly 0 for all $\Gamma$, $\nu_{obs}$ and $E_{p,obs}$ values with $q_{obs}=0.5$.

   \begin{figure}
   \centering
   \includegraphics[width=\textwidth, angle=0]{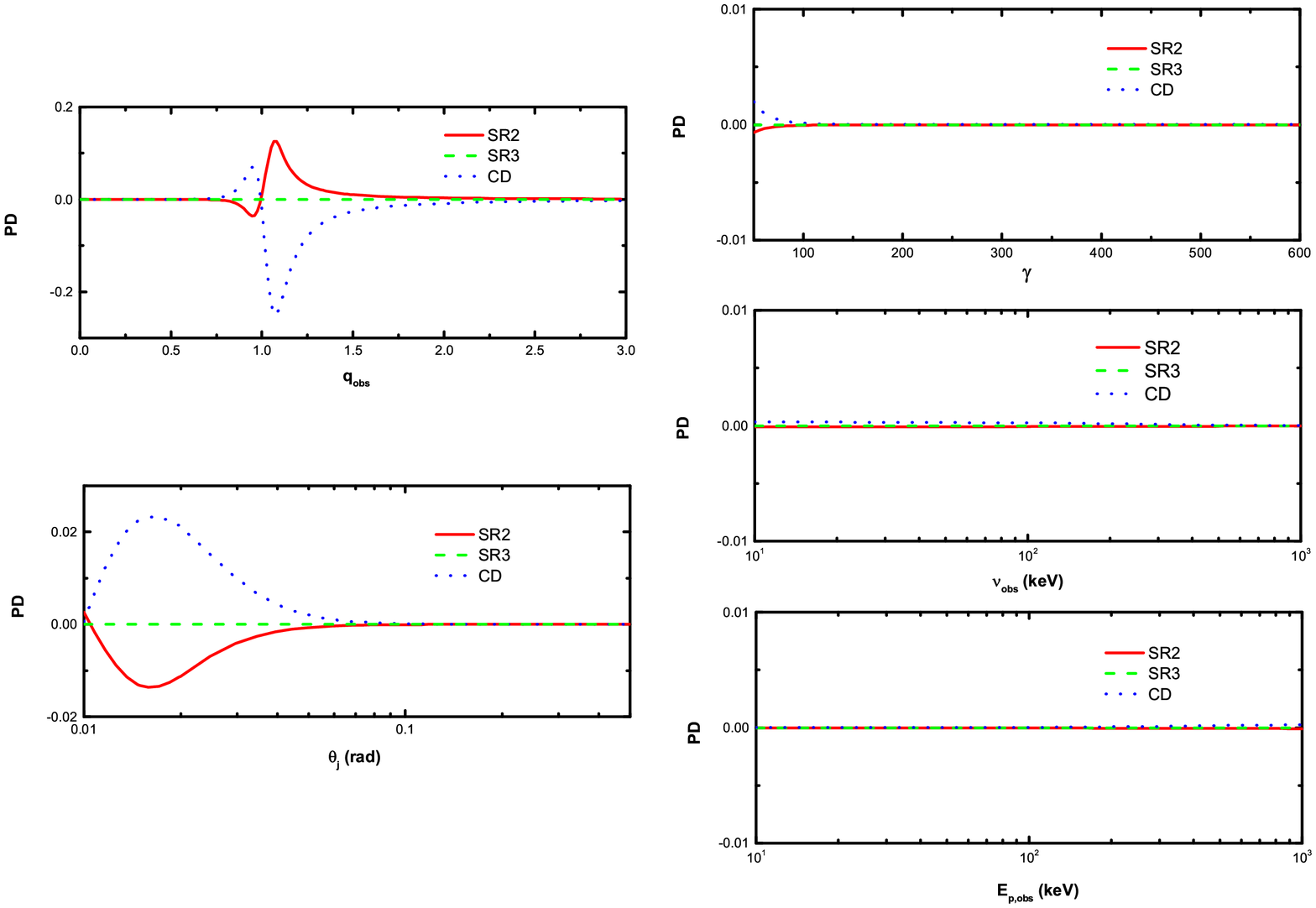}
   \caption{PD evolutions with $q_{obs}$, $\theta_j$, $\Gamma$, $\nu_{obs}$ and $E_{p,obs}$ for SR2, SR3 and CD models.}
   \label{Fig1}
   \end{figure}

The polarization properties of SM models evolving with $\xi_B$ are shown in Fig. 8. For SMR model, even for SOR model (i.e., $\xi_B\rightarrow\infty$), its PD value is roughly 0 for the observational geometry of $\theta_j=0.1$ and $q_{obs}=0.5$, which is consistent with that shown in Fig. 2. PD curves of SMA and SMT models coincide with each other. The fast rise phases of these two curves continue till $\xi_B=3$ and then they keep roughly as constant. When $\xi_B=0$, PDs for three SM models are 0 because the total magnetic field is random and isotropic in 3D space (Lan et al. 2019). For SMA and SMT models, when $\xi_B\gg1$ (i.e., $\xi_B\geq3$), PDs for these two SM models will approach $\sim25\%$, of which the corresponding SO models can reached. PA for $\xi_B=0$ of SMA model is meaningless, because when $\xi_B=0$ there is no net polarization and the two Stokes parameters $Q_{\nu}$ and $U_{\nu}$ are both zero. Because of the computational error of the computer $Q_{\nu}$ and $U_{\nu}$ for $\xi_B=0$ are very tiny but non-zero, leading to the ``computational" PA for $\xi_B=0$.

   \begin{figure}
   \centering
   \includegraphics[width=\textwidth, angle=0]{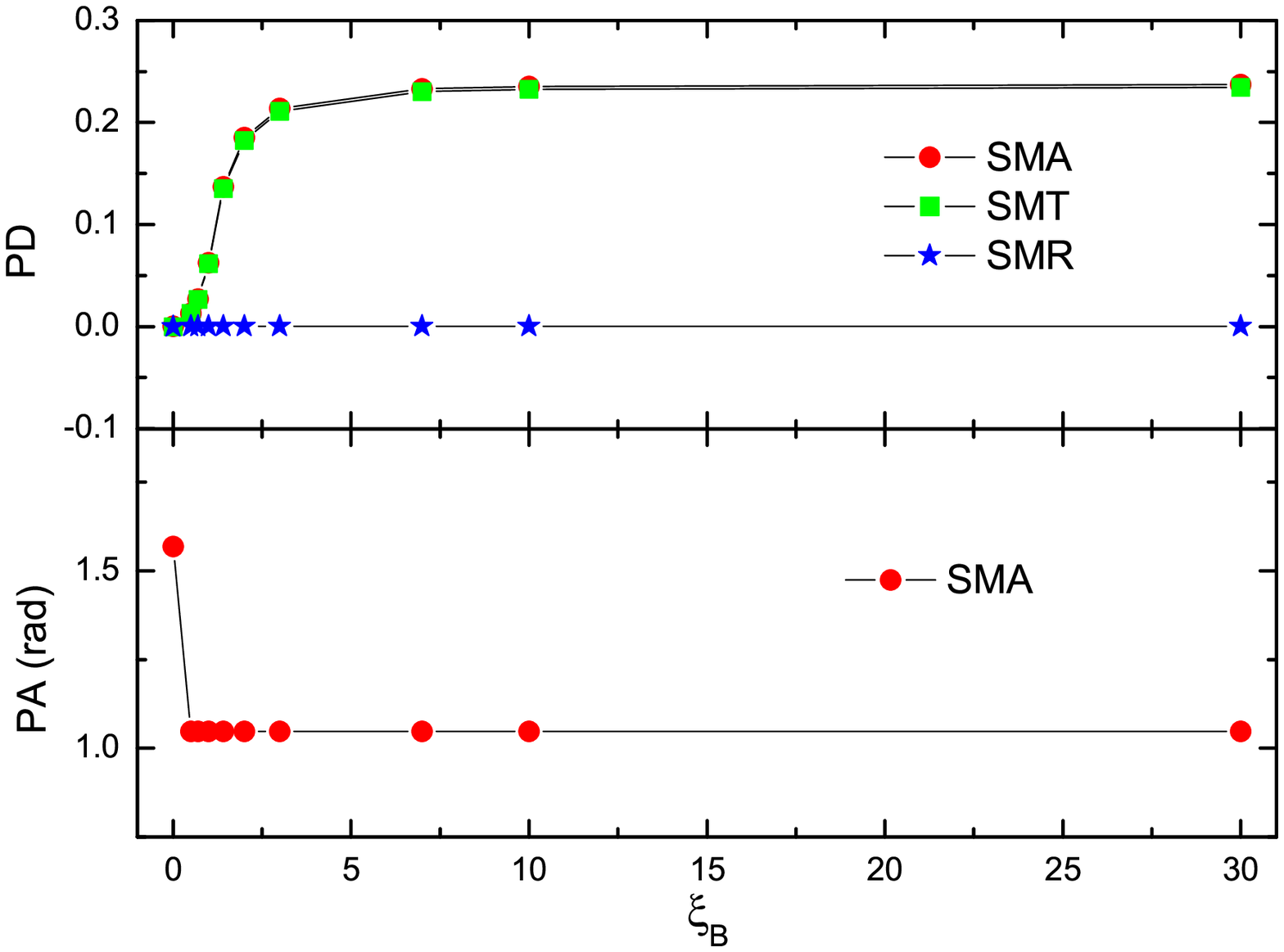}
   \caption{Polarization evolutions with different $\xi_B$ values for the SMA (red circles), SMT (green diamonds) and SMR (blue stars) models. The upper panel shows the PD curves for three SM models and the lower panel shows the PA curve for the SMA model. The points are our numerical data.}
   \label{Fig1}
   \end{figure}

\section{Statistics}
\label{sect:analysis}
In this section, Monte Carlo simulations have been performed and the statistical properties of polarization of GRB prompt emission have been studied.
Since the polarization properties of the 9 models discussed in this paper are not very sensitive to the observational frequency and also the frequency-integrated polarization is very expensive for computing resource, we will consider the polarization properties at single frequency in the following statistical study. We have simulated random numbers for $q_{obs}$, $\theta_j$, $z$, $E_{iso}$ and $E_p$. $E_{iso}$ is the isotropic equivalent energy in $\Gamma-$ray band of a GRB, $E_p=E_{p,obs}(1+z)$ is the peak energy in the burst source frame. The probability density function (PDF) of redshift is assumed to be proportional to star formation rate (Porciani \& Madau 2001).
\begin{equation}
f_1(z)=\frac{e^{3.4z}}{e^{3.8z}+45}\frac{\left(\int dz_1/\sqrt{\Omega_M(1+z_1)^3+\Omega_\Lambda}\right)^2}{(1+z)^{3/2}}
\end{equation}
where $\Omega_M=0.3$ and $\Omega_\Lambda=0.7$ are the normalized density for matter and dark energy, respectively. The PDF ($f_2(\theta_j)$) of the half-opening angle of GRB jet is taken from Fig. 4(a) of Goldstein et al. (2016), which is derived through Ghirlanda relation (Ghirlanda et al. 2004) and is suitable for long GRBs. We take the fluence-corrected PDF for $q_{obs}$ (Gill et al. 2018).
\begin{equation}
f_3(q_{obs})=\bar{f}_{iso}(q_{obs})q_{obs}=\left[\int^{\theta_{j,max}}_{\theta_{j,min}}\tilde{f}_{iso}(\theta_V,\theta_j)f_2(\theta_j)d\theta_j\right]q_{obs}
\end{equation}
$\tilde{f}_{iso}(\theta_V,\theta_j)=E_{iso}(\theta_V,\theta_j)/E_{iso}(0,\theta_j)$ (Salafia et al. 2015), which is also the ratio of fluence at $\theta_V$ to that at $\theta_V=0$.
\begin{equation}
\tilde{f}_{iso}(\theta_V,\theta_j)=\frac{\int^{\theta_j}_0\sin\theta d\theta\int^{2\pi}_0\frac{d\phi}{(1-\beta\cos\alpha)^3}}{\frac{\pi}{\beta}\left(\frac{1}{(1-\beta)^2}-\frac{1}{(1-\beta\cos\theta_j)^2}\right)}
\end{equation}
where $\cos\alpha=\cos\theta\cos\theta_V+\sin\theta\sin\phi\sin\theta_V$. Then we marginalize $\theta_j$ using its PDF to get the fluence-weighted factor $\bar{f}_{iso}(q_{obs})$. For the top-hat jet discussed in this paper, the PDF of $q_{obs}$ drops quickly when it is larger than 1, therefore, most of our simulated GRBs are detectable.
The random numbers of $E_{iso}$ and $E_p$ are generated through empirical relations, which reads $E_{iso}\theta^2_j/2=10^{51}\zeta_1$ erg and $E_p=200\zeta_2(E_{iso}/10^{52}$ erg$)^{1/2}$ keV, where $\zeta_1$ and $\zeta_2$ are assumed to obey the lognormal distribution, the averages of these two random numbers are set to be 1 and the logarithmic variance of $\zeta_1$ and $\zeta_2$ are 0.3 and 0.15, respectively (Toma et al. 2009).

To calculate the exact fluence, we need $R^2A_0$ and $\nu'_0$. For an on-axis observer, we have $E_p(\theta_V=0,\theta_j)\doteq\bar{\mathcal{D}}\nu'_0$ and $E_{iso}(\theta_V=0,\theta_j)\doteq(16\pi^2/e)(E_p/h)R^2A_0\Gamma^2\theta_j^2/(1+\Gamma^2\theta_j^2)$, where e is the base number of nature logarithm and $\bar{\mathcal{D}}=\int^{\theta_j}_0\mathcal{D}\sin\theta d\theta/\int^{\theta_j}_0\sin\theta d\theta\doteq2\ln(1+\Gamma^2\theta_j^2)/(\Gamma\theta_j^2)$. In deriving the expressions for $\bar{\mathcal{D}}$ and $E_{iso}(\theta_V=0,\theta_j)$, the approximation $\Gamma\gg1$ and $\theta_j\ll1$ are used. But for each set of random numbers, $\theta_V$ is rarely to be 0, we need to transform $E_{iso}(\theta_V,\theta_j)$ and $E_p(\theta_V,\theta_j)$ to the corresponding on-axis qualities through $E_{iso}(\theta_V=0,\theta_j)=E_{iso}(\theta_V,\theta_j)/\tilde{f}_{iso}(\theta_V,\theta_j)$ and $E_p(0,\theta_j)=E_p(\theta_V,\theta_j)/\tilde{f}_{iso}^{1/2}(\theta_V,\theta_j)$.

We then calculate the statistical properties for SO, SR2 and CD models. Except for the simulated random numbers, the other parameters used in statistical calculation are  $\alpha_s=-0.2$, $\beta_s=1.2$ and $\Gamma=100$. For SOA model, the orientation of the aligned magnetic field is assumed to be $\delta_a=\pi/6$. The observational frequency is taken as $h\nu_{obs}=250$ keV. Fig. 9 shows our simulated results for SOA and SOT models. For these two models, there is a PD island in $E_{p,obs}-PD$ diagram, the PD value at this PD island is about $25\%$ and PDs of most simulated GRBs take this value. PD distributions of the simulated GRBs in $q_{obs}-PD$ diagram shown in Fig. 9 trace the corresponding curves in Fig. 1 with scatters due to the distribution of parameters. This result is not a coincidence, because except $\theta_V$, we take fixed parameters for others in Fig. 1. The fixed values of parameters used in this paper are often the values with maximum possibility in our simulation. Therefore, the simulation results can be inferred from the $q_{obs}-PD$ curves in Figs. 1 and 7, i.e., PDs of the simulated GRBs in $E_{p,obs}-PD$ diagram will concentrate around the PD value at PD plateau in $q_{obs}-PD$ curve when $q_{obs}<1$. There are some GRBs for SOT model laying at the left lower corner of $q_{obs}-PD$ diagram. The $q_{obs}$ parameters of these GRBs are small and so do their PDs. The smaller $q_{obs}$ parameter indicates that these GRBs are viewed nearly on-axis hence are very bright. The lower PDs for nearly on-axis observations would prefer the SOT model. Therefore, we conclude that bright GRBs with lower PDs than that of PD value at PD island would favour the SOT (or SMT) model.

   \begin{figure}
   \centering
   \includegraphics[width=\textwidth, angle=0]{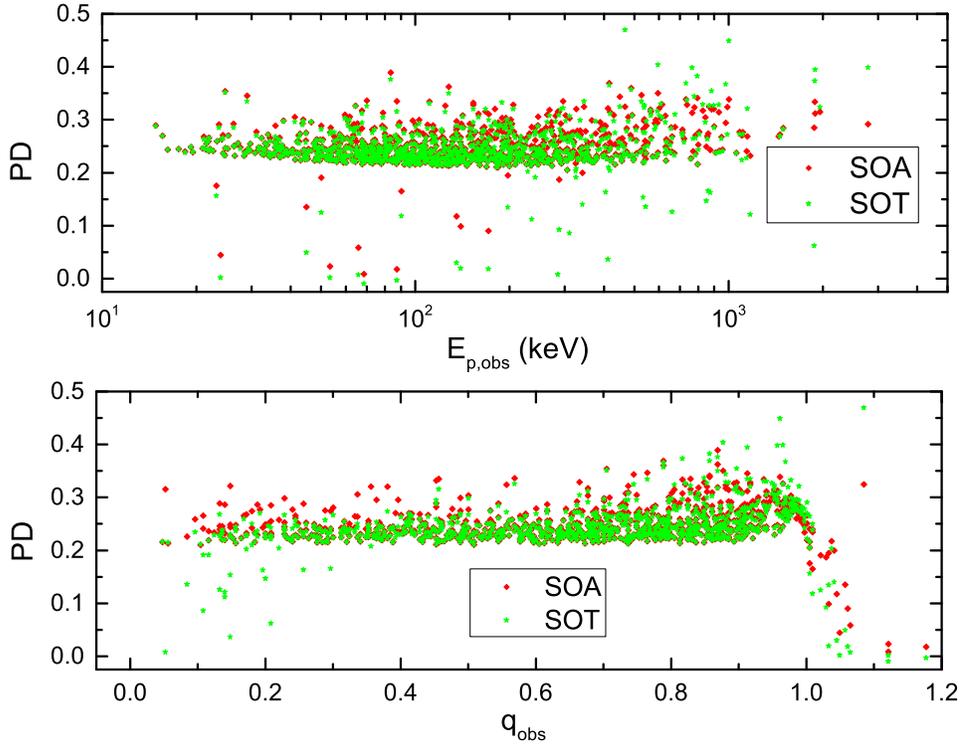}
   \caption{PD distributions of the simulated GRBs for SOA and SOT models. The upper and lower panels show the $E_{p,obs}-PD$ and $q_{obs}-PD$ diagrams, respectively.}
   \label{Fig1}
   \end{figure}

The $E_{p,obs}-PD$ diagrams of SOR, SR2 and CD models are shown in Fig. 10. There is a PD island for these three models with PD $\sim$ 0 and PD value at that PD island in $E_{p,obs}-PD$ diagram equals to that in PD plateau of $q_{obs}-PD$ curves in Figs. 1 and 7 with $q_{obs}<1$. Following the conclusions of SO, SR2 and CD models, we do not simulate the PD distribution for SM models. It can be inferred from Fig. 1 that PDs of both SMA and SMT models will also concentrate around some value and this concrete PD value depends on $\xi_B$ parameter, i.e., there will be a PD island in $E_{p,obs}-PD$ diagram for SMA and SMT models and the PD value at this PD island can range from 0 to $25\%$. Here, the distributions of PA for various models are not discussed, because the orientations of GRB jets in the sky are different, leading to the reference system of PA will vary from burst to burst.

   \begin{figure}
   \centering
   \includegraphics[width=\textwidth, angle=0]{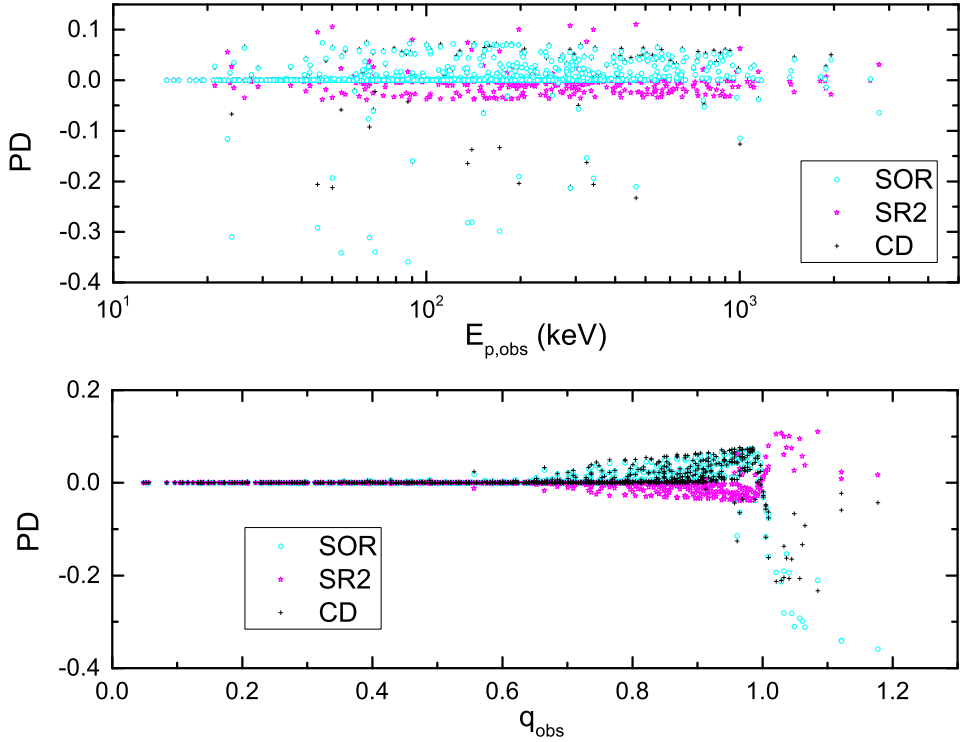}
   \caption{Same as Fig. 9, but for SOR, SR2 and CD models.}
   \label{Fig1}
   \end{figure}

Fig. 6 of Toma et al. (2009) had shown the statistical properties for SOT, SR2 and CD models. For SR2 and CD models, PDs of our simulated GRBs can also be negative depending on the sign of Stokes parameter $Q_\nu$, because of our calculation formula defined in Sec. 2. Then the absolute value of PD represent the amplitude of polarization and its sign shows its polarization direction. In Toma et al. (2009), they take the absolute value of $Q_\nu$ in their calculation hence lead to a positive PD. The information of polarization direction will be lost in their treatment. Polarization direction will be unimportant in the statistical study because the orientation of each GRB jet would be different in the sky leading to incomparability of PAs. In Fig. 6 of Toma et al. (2009), PD values at PD islands for SR2 and CD models are also 0 and our results are consistent with their results. For SOT model, PD value at PD island is about $\sim35\%$ in Toma et al. (2009), while it is $\sim25\%$ in our simulation and the reason for this will be discussed in the following Conclusion and Discussion section.

\section{Application to POLAR's Observations}
\subsection{$\sim10\%$ time-integrated PD}
POLAR is a $\gamma$-ray polarization detector, which is onboard Tiangong-2 space laboratory of China. During its operation, high-quality polarization observation had been made for 5 GRBs (Zhang et al. 2019). Interestingly, the time-integrated PDs of these measured bursts are around $10\%$. While PD lower limit of former observations for GRB prompt emission is about $30\%$, which means that GRB prompt emission is highly polarized and SO models (i.e., SOA and SOT models) are favoured. Although the PD ``upper limit" of POLAR's data is consistent with that of the former observations, its results do show that most of the GRBs may be moderately polarized. Since POLAR has detected polarization properties of 5 GRBs, these results are meaningful for statistical study.

PDs of the observed bursts are concentrated around $10\%$, which is very likely to be PD value at PD island of $E_{p,obs}-PD$ diagram. That value usually equals to PD value at PD plateau when $q_{obs}<1$ of $q_{obs}-PD$ curve. In Sec. 4, PD islands of SO, SR and CD model are either too high ($\sim25\%$) or too low ($\sim0\%$) compared with the observed $\sim10\%$ PD. Since PDs of a SM model can range from $0\%$ to that of a corresponding SO model ($\sim25\%$), the observed $\sim10\%$ PD will favour SMA and SMT models. \footnote{For SMR model, its PD will concentrated around 0, hence is disfavour.} From theoretical aspect, magnetic field of internal shock model or magnetic reconnection model would be mixed, i.e., the ordered magnetic field carried out form the central engine will be disturbed by collisions, shock or magnetic reconnection. Our result of a mixed magnetic field in the emission region of GRB prompt phase agrees well with the prediction of popular models. Because PD value at PD plateau of $q_{obs}-PD$ curve for SMA and SMT models depends strongly on the $\xi_B$ parameter, if we take typical values for other parameters, $\xi_B$ of these observed GRBs is constrained to be 1.135.

\subsection{GRB 170114A}
GRB 170114A is very bright, so observations, at least for the peak of the light curve, are very likely to be on-axis. Time-resolved PDs of this burst seem to be large (about $30\%$ around the light curve peak), so SR2, SOR, SMR and CD models are disfavoured, because large PD will be obtained only for off-axis observation of these two models. PA of GRB 170114A can evolve both gradually and abruptly by $\sim90^\circ$. For the non-precessing jet, especially for the one-emission-region models, abrupt $90^\circ$ PA change is very rare for SO and SM models, which can be seen from Figs. 1, 3-6 and 7, hence these models are also disfavoured. Recently, polarization properties involving a precessing jet had been discussed by Lan et al. (2019). PA of a precessing jet can evolve both gradually and abruptly by $\sim90^\circ$ for both SR2 and SOA models. Since SR2 model is rejected by high time-resolved PD of this burst, we only consider SOA model with a precessing jet. The time-integrated PD of the burst is relatively small ($\sim4\%$), compared with that of time-resolved PD. For SOA model of a precessing jet, since the abrupt $90^\circ$ PA change is very rare, the cancellation of time-resolved polarized flux will be not significant and the resultant time-integrated PD will not reduce much compared with that of time-resolved PD. Therefore, SOA model with a precessing jet is also disfavoured. For the magnetic patch model (Granot \& K\"{o}nigl 2003; Nakar \& Oren 2004; Granot \& Taylor 2005), its PA will evolve randomly. Since PA of GRB 170114A seems to be rotating anti-clockwise with time, this model is also disfavoured. Finally, it seems that no polarization model on hand can explain the observations of GRB 170114A.

\section{The polarization of GRB 110721A}
The PD of GRB 110721A is very high and
the best fit value reaches $84\%$. Polarization of the burst was interpreted with the
early reverse shock model (Fraija et al. 2017a). Such a high PD is
even larger than that of synchrotron emission of power-law electrons in an
ordered magnetic field (i.e., (p+1)/(p+3/7) with p the power-law index of electron
spectrum). For the models discussed in this paper, a power-law distributed electron spectrum is
adopted. PD of the jet emission predicted by the
models with an ordered magnetic field is at a level of $25\%$ (see our Fig. 9)
and the maximum predicted PD of the jet emission is $<50\%$, which is smaller than the
lower limit ($\sim 56\%$) of GRB 110721A. Therefore, it seems that PD predictions of the models in this paper
could not interpret the PD observations of GRB 110721A. However, the detection confidence level of GRB 110721A is relatively low.
PD can be less than $20\%$ or as high as $100\%$ for 3$\sigma$
confidence level. Actually, Up till now, there has not been a confirmation polarization
detection with a 5$\sigma$ confidence level made in GRB prompt phase. Because
of high uncertainty of the the observational data, it is hard to definitely discriminate
the theoretical models.

MFCs during the internal shock phase are also uncertain. Even
the shock generated magnetic field might be random (Gruzinov \& Waxman 1999; Inoue et al. 2011),
it is still possible that a large-scale ordered magnetic field
advected from the GRB central engine can survive during the internal shock phase. In Fan et al. (2004),
the internal shocks with magnetization was discussed. Polarization with such a magnetized internal shock
was also calculated roughly and the maximum PD predicted by the model is 0.6, which is the PD value
in a purely ordered magnetic field. Actually, the maximum PD ($\sim 0.6$) adopted in Fan et al. (2004) is very conservative.
Depending on the MFC and energy spectrum of electrons, PD of the jet emission with an ordered
magnetic field can reach as high as $\sim 0.9$ (Lan \& Dai 2020). Therefore, in principle, PD observations
of GRB 110721A cannot reject the magnetized internal shock model.

Our model cannot discriminate between the internal shocks and reverse
shocks for GRB 110721A. Because the model in our paper is constructed from
observations of the GRB prompt energy spectrum and it is independent of the
internal shock model and of the reverse shock model. Therefore, we cannot use
the results of our model to discriminate these two models.

The major factors that affect
the GRB polarization are the MFCs and energy spectra of electrons
in the emission region. For these two models (internal shock and reverse shock),
the emission matters are both ejected from the central engine, so it is possible
for these two models to have a large-scale ordered magnetic field in their emitting
region. Electrons in these two models are both mainly accelerated by shocks, and
energy spectra of shocked accelerated electrons might be similar. Therefore, polarization
predictions of these two models should be similar. The predicted PD of these two
models can range from zero (for 3D random magnetic field) to $\sim 90\%$ (for single-energy
electrons in an ordered magnetic field, see Lan \& Dai 2020). The predicted PAs of the most models are usually
a constant. The abrupt $90^\circ$ and the gradual PA changes are relatively rare.

\section{The polarization of prompt optical flash of GRB 160625B}

Troja et al. (2017)
have reported their polarization observations of prompt optical flash of GRB 160625B.
The observed PD is variable and significant, ranging from $\sim 5\%$ to $\sim 8\%$.
In GRBs, the main emission mechanism is synchrotron and then three factors affect
polarization properties significantly, the MFC, jet structure and
observational geometry. Fraija et al. (2017b) have modeled the early and late afterglows
of the burst and they found that the ejecta of GRB 160625B is magnetized, with a
magnetization parameter of $\sigma\simeq0.4$. The energy spectra of GRB 160625B
were analysed by Zhang et al. (2018) and the emission region of the main burst was
found to be Poynting-flux dominated. These results can be confirmed by the polarization
observations of prompt optical flash of the burst. The optical linear PD
increases from about $5\%$ to $8\%$. And GRB 160625B is very bright, indicating an on-axis
observation. Combining these two facts, the favoured MFC
in the emitting region of optical radiation is mixed, i.e., including both ordered and
random components (Lan et al. 2019). Because PD with a random magnetic
field is about zero for an on-axis observation, while it will be $\sim25\%$ for a purely ordered
magnetic field (see Figs. 9 and 10 in this paper). The mixed magnetic field in the optical emitting region means the
ordered part of magnetic field exists in the ejecta and will indicate a magnetized ejecta
from central engine.

As mentioned in Section 6, the results of the models in this paper
cannot be used to distinguish the internal shocks and reverse shocks. By assuming the
synchrotron emission, the model in this paper is constructed from observations of GRB
prompt phase. And it is only suitable for the emitting spectrum with a Band-function like
form. Therefore, whether the model in this paper can be used to describe the optical
polarization of GRB 160625B or not, depends on the energy spectrum of the optical flash.
Actually, the two models have similar properties
that affect polarization significantly. The materials of both internal shocks and reverse
shocks are ejected from the central engine. If the outflow is magnetized, there will be
large-scale ordered magnetic field (or at least mixed magnetic field) in these materials,
large PD (or at least moderate PD) can be predicted for these two models. Polarization
predictions of these two model should be similar. If the MFCs in these two models are mixed,
both models could predict the observed $5\%$ to $8\%$ PD of the optical flash of GRB 160625B.
As mentioned above, the observed
moderate PDs of the optical flash also suggest a magnetized ejecta, while we do not
know that the emission of prompt optical flash of GRB 160625B is whether from the
magnetized internal shocks or from the reverse shocks. Therefore, it might be very hard
to distinguish these two models through polarization observations.

\section{Conclusions and Discussion}

Because polarizations of synchrotron emission are very sensitive to MFCs, their properties with different kinds of MFCs should be investigated in detail. In this paper, we have discussed the polarization properties of GRBs with four new models (i.e., SMA, SMT, SMR and SR3) and compared their properties with these of SOA, SOT, SR2 and CD models. Then a set of random numbers has been simulated and the statistic properties of GRB polarization are studied.

In SOA model, the aligned magnetic field is assumed to be a series of parallel lines in the plane of sky, while the ordered part is latitude circles in the jet surface in SMA model. Through our calculation, the polarization properties of SOA and SMA models are indeed very similar, which infers that the difference of aligned magnetic field in SOA and SMA models is very tiny, parallel lines in the plane of sky are good approximation of latitude circles in the jet surface. In our treatment, SR3, SMR and SOR models are handled separately and these results are consistent. Polarization properties of SMR model approaches that of SR3 model when $\xi_B\rightarrow0$ and approaches that for SOR model when $\xi_B\gg1$.

Polarization properties of SM models and the corresponding SO models (e.g. SMA and SOA) are very similar, except that PD of the SM models can be lower, depending on $\xi_B$ values. PDs of SO and SM models are sensitive to the observational angle $\theta_V$, jet half-opening angle $\theta_j$ and peak energy $E_{p,obs}$, but are insensitive to the bulk Lorentz factor $\Gamma$ and observational frequencies $\nu_{obs}$. The conclusion for $\nu_{obs}-PD$ dependence is suitable only for the energy band between 10 and 1000 keV. PA evolutions are rare. For SMA and SOA models, PAs change gradually only when the $1/\Gamma$ cone crosses the jet cone. PD plateau of SOT model in $q_{obs}-PD$ curve of Fig. 1 when $q_{obs}<1$ is about $25\%$ in our calculation, while it is $\sim40\%$ in Toma et al. (2009), because $E_{p,obs}$ is taken as 350 keV in our calculation while it is roughly 16 keV in Toma et al. (2019).\footnote{In Toma et al. (2009), $\Gamma h\nu'_0=350$ keV and $\Gamma=100$, then $h\nu'_0=3.5$ keV, and $E_{p,obs}=E_p(\theta_V,\theta_j)/(1+z)=\tilde{f}_{iso}^{1/2}(\theta_V,\theta_j)\bar{\mathcal{D}}\nu'_0/(1+z)\sim16$ kev.} From our Fig. 10, PD is about $25\%$ when $E_{p,obs}=350$ keV, while it is as high as $40\%$ for $E_{p,obs}=16$ keV. Therefore, our results of SOT model are consistent with that of Toma et al. (2009).

It is known that polarization originates from asymmetries. The change of PD will reflect the change the asymmetry in the system, including that of the emission region itself and of the location of the observer. When the asymmetry of the system increases, PD also increase and vice versa. Generally speaking, there are two kinds of asymmetries for synchrotron emission in GRB jet, one originates from the magnetic field in the emission region and the other is contributed by the geometry (including the geometry of jet structure and of observation). The evolutions of PD value can be analysed by the changes of these asymmetries in the system. In addition, the spectral index of electrons also affect local PD significantly and then the PD of the jet.

$q_{obs}-PD$ diagrams of the simulated bursts trace the corresponding $q_{obs}-PD$ curves in Figs. 1 and 7 with small scatters. PD values at PD islands in $E_{p,obs}-PD$ diagrams are also the PD values of PD plateaux in $q_{obs}-PD$ curves with $q_{obs}<1$ of Figs. 1 and 7. PD islands in $E_{p,obs}-PD$ diagrams of SOA and SOT models are both concentrated around $25\%$, so it is hard to distinguish these two models through the statistics of PD values. Same conclusions can be made for SMA and SMT models, because polarization properties of SMA (SMT) model are very similar to that of SOA (SOT) model, which can be seen from Fig. 1. PD values at PD islands in $E_{p,obs}-PD$ diagrams of both SMA and SMT models can range from 0 to $25\%$, depending on $\xi_B$ values. PDs for SR3 model are always 0, independent of parameters. Although the maximum PDs reached by CD and SOR models are higher, it is still hard to distinguish them from SMR and SR2 models through statistics. Because the locations and orientations of the jet axes of the simulated bursts can be different, it is meaningless to discuss the statistical properties of PA.

It was shown in the former studies that PA can evolve gradually for SOA (SMA) model, while it can only change abruptly by $90^\circ$ for SOT (SMT) model, then the two models are distinguishable through PA evolution patterns of the single burst (Lan et al. 2016a,2019). Here, we suggest that SOA (SMA) and SOT (SMT) models can also be distinguishable through statistics of their PDs. If we have a large time-integrated PD sample of GRB prompt phase, there is a non-zero PD island for these GRBs in the sample, then SOA, SOT, SMA and SMT models are favoured. Selecting the bright bursts (which is very likely to be observed on-axis), then if PDs of at least several bright GRBs are substantially lower than PD value at PD island, SOT (SMT) models will be favoured. If PDs of all the bright GRBs are around the PD island value, SOA (SMA) model are favoured. If the PD island value of observed sample is around 0, then SOR, SR2 and CD models are favoured.

Finally, we apply our simulation results to POLAR's data and find that SMA and SMT models are mostly favoured for the observed time-integrated $\sim10\%$ PD. For GRB 170114A, large time-resolved PD favors the SO models. Both the gradual and abrupt $\sim90^\circ$ PA changes of the burst favor the SOA model with a precessing jet, while a low $\sim4\%$ time-integrated PD could not be obtained from the SOA model of a precessing jet, hence the model is disfavoured. The magnetic patch model is also disfavoured by the roughly anti-clockwise rotated PA of GRB 170114A.

\begin{acknowledgements}
We thank Hai Yu and Shuang-Xi Yi for useful discussions and also thank En-Wei Liang for helpful comment. This work is supported by the National Key Research and Development Program of China (grant no. 2017YFA0402600) and the National Natural Science Foundation of China (grant no. 11573014, 11673068, 11725314, 11833003 and 11903014). X.F.W. is also partially supported by the Key Research Program of Frontier Sciences (QYZDB-SSW-SYS005), the Strategic Priority Research Program ``Multi-waveband gravitational wave Universe'' (grant No. XDB23040000) of the Chinese Academy of Sciences and the ``333 Project" of Jiangsu province. M.X.L is supported by the Natural Science Foundation of Jiangsu Province (grant No. BK20171109) and by the Fundamental Research Funds for the Central Universities, in part by National Science Foundation of China (NSFC) under Grant No. 11847310 and the Seeds Funding of Jilin University.
\end{acknowledgements}

\appendix                  

\section{Polarization of Synchrotron Emission with Different MFCs and CD Model}

\subsection{Synchrotron Emission in an Ordered Magnetic Field (SO)}
An ordered magnetic field in the emission region is still possible, of which can be carried out from GRB central engine. Since the directions of the ordered magnetic fields are fixed in a point-like region (at which the direction of the comoving wavevector is roughly fixed), the local PD $\Pi_p$ for these three models, i.e., SOA, SOT and SOR, will be equal to $\Pi_0$, where $\Pi_0=(\tilde{\alpha}+1)/(\tilde{\alpha}+5/3)$ is the PD of the synchrotron emission in an ordered magnetic field. And $A_0$ for three models can be expressed as $A_0=(\sin\theta'_B)^{\tilde{\alpha}+1}$. The expressions of $\sin\theta'_B$ and local PA $\chi_p$ for SOA model are shown successively in the following (Lan, Wu \& Dai 2016a).
\begin{equation}
\sin\theta'_B=\left[1-D^2\frac{\sin^2\theta\cos^2(\phi-\delta_a)}{\cos^2\theta+\sin^2\theta\cos^2(\phi-\delta_a)}\right]^{1/2},
\end{equation}
\begin{equation}
\chi_p=\phi+\arctan\left(\frac{\cos\theta-\beta}{\cos\theta(1-\beta\cos\theta)}\cot(\phi-\delta_a)\right),
\end{equation}
where $\delta_a$ is the orientation of the aligend magnetic field. These formulas for SOT model are as follows (Toma et al. 2009; Lan, Wu \& Dai 2016a)
\begin{equation}
\sin\theta'_B=\left[1-D^2\frac{\sin^2\theta_V\sin^2\theta\sin^2\phi}{\sin^2\theta\sin^2\phi+(\sin\theta_V\cos\theta-
\cos\theta_V\sin\theta\cos\phi)^2}\right]^{1/2},
\end{equation}
\begin{equation}
\chi_p=\phi+\arctan\left(\frac{\cos\theta-\beta}{\cos\theta(1-\beta\cos\theta)}\times \frac{\sin\theta_V\cos\theta\sin\phi}{(\cos\theta_V\sin\theta-\sin\theta_V\cos\theta\cos\phi)}\right).
\end{equation}

In the following, we will derive the expressions of $\sin\theta'_B$ and of the local PA $\chi_p$ for the SOR model. In this model, the magnetic field is along the radial direction and also we assume that jet has no lateral expansion, which finally reads $\hat{B}'=\hat{\beta}$, where $\hat{\beta}$ is the velocity direction of the local fluid element. Then the electric vector of synchrotron photons is $\hat{e}\parallel\hat{\beta}\times\hat{k}$, where $\hat{k}$ is the wavevector in the observer frame. Then we establish a global coordinate system $\hat{X}\hat{Y}\hat{k}$, with $\hat{X}$ along the projection of the jet axis in the plane of sky. The polar and azimuthal angle of the local velocity $\hat{\beta}$ in $\hat{X}\hat{Y}\hat{k}$ system are $\theta$ and $\phi$, then $\hat{\beta}=(\sin\theta\cos\phi,\sin\theta\sin\phi,\cos\theta)$. After some calculations, we finally get $\hat{e}=\sin\phi\hat{X}-\cos\phi\hat{Y}$. Then the local PA for SOR model is
\begin{equation}
\chi_p=\arctan\left(\frac{e_Y}{e_X}\right)=\phi+\frac{3\pi}{2},
\end{equation}
And the pitch angle of electrons in such a radial magnetic field can be found through $\cos\theta'_B=\hat{B}'\cdot\hat{k}'=(\cos\theta-\beta)/(1-\beta\cos\theta)$, where $\hat{k}'$ is the comoving wavevector.
\begin{equation}
\sin\theta'_B=\sqrt{1-\frac{(\cos\theta-\beta)^2}{(1-\beta\cos\theta)^2}}
\end{equation}

\subsection{Synchrotron Emission in a Mixed Magnetic Field (SM)}
During jet propagation, collisions, shocks or magnetic reconnections may happen, which will disturb the magnetic field lines, leading to a mixed magnetic field. Here we also consider three kinds of mixed magnetic fields with different ordered components (i.e., SMA, SMT and SMR), which is same as that in Lan et al. (2019). These three ordered magnetic field components in the mixed magnetic fields are same as that discussed above in Sec. 2.1. The aligned and toroidal ordered components are assumed to be confined in the shock plane while the radial ordered component is along the radial direction of the jet element. The random part of the mixed magnetic field is assumed to be isotropic in 3 dimensional space.

Same as that in Lan et al. (2018), we establish two coordinate systems: $\hat{x}\hat{y}\hat{\beta}$ and $\hat{1}\hat{2}\hat{k}'$, where $\hat{y}=\hat{1}\parallel\hat{\beta}\times\hat{k}$. In a smaller region, where the direction of the magnetic field is fixed, let the polar and azimuthal angles of the magnetic field in $\hat{1}\hat{2}\hat{k}'$ system be $\theta'_B$ and $\phi'_B$. The detailed derivations of the local PD and of the local PA are not repeated here, which can be found in Lan et al. (2019). We only give the final results, which will be used here.
\begin{equation}
\Pi_p=\Pi_0\frac{\sqrt{\langle(\sin\theta'_B)^{1+\tilde{\alpha}}\cos(2\phi'_B)\rangle^2+\langle (\sin\theta'_B)^{1+\tilde{\alpha}}\sin(2\phi'_B)\rangle^2}}{\langle(\sin\theta'_B)^{1+\tilde{\alpha}}\rangle},
\end{equation}
\begin{equation}
\chi_p=\phi+\chi'_p+\frac{\pi}{2}+n\pi
\end{equation}
with
\begin{equation}
\chi'_p=\frac{1}{2}\arctan\left(\frac{\langle (\sin\theta'_B)^{1+\tilde{\alpha}}\sin(2\phi'_B)\rangle}{\langle (\sin\theta'_B)^{1+\tilde{\alpha}}\cos(2\phi'_B)\rangle}\right)
\end{equation}
The angle bracket denotes the average over the magnetic field direction. $A_0$ can be expressed as $A_0=\langle (\sin\theta'_B)^{\tilde{\alpha}+1}\rangle$. The average over the magnetic field direction and the expressions for $\sin\theta'_B$, $\sin\phi'_B$ and $\cos\phi'_B$ for three kinds of mixed magnetic field with different ordered components can also be found in Lan et al. (2019). $n$ is an integer.

\subsection{Synchrotron Emission in a Random Magnetic Field (SR)}
The random magnetic field may be generated or amplified during the shock propagation, which is favoured by the observed low PD values (a few\%) during the late GRB afterglow phase (Covino et al. 1999; Rol et al. 2000,2003; Gorosabel et al. 2004; Greiner et al. 2004; Wiersema et al. 2012). Literally, an anisotropic 3D random field had been discussed by several authors (e.g., Sari 1999; Gruzinov 1999). Here, both SR2 (Toma et al. 2009; Lan et al. 2019) and SR3 models are considered. For these two models, we have $A_0=\langle (\sin\theta'_B)^{\tilde{\alpha}+1}\rangle$ and $\Pi_p=|\langle Q'_p\rangle/\langle F'_p\rangle|=|-\Pi_0\langle (\sin\theta'_B)^{\tilde{\alpha}+1}\cos(2\phi'_B)\rangle/\langle (\sin\theta'_B)^{\tilde{\alpha}+1}\rangle|$.

For SR2 model, consider a smaller region, where the magnetic field direction is fixed, because the 2 dimensional random magnetic field is assumed to be confined in the shock plane, $\eta'$ is set be the angle between the magnetic field and $x$-axis in $\hat{x}\hat{y}\hat{\beta}$ coordinate. Then the 2 dimensional random magnetic field confined in the shock plane can be expressed as $\hat{B}'=\hat{B}'_2=\cos\eta'\hat{x}+\sin\eta'\hat{y}$. The expressions for $\sin\theta'_B$ and $\cos(2\phi'_B)$ of SR2 model can be found in Toma et al. (2009) and Lan, Wu \& Dai (2016), which reads
\begin{equation}
\sin\theta'_B=\left(1-D^2\sin^2\theta\cos^2\eta'\right)^{1/2},
\end{equation}
\begin{equation}
\cos(2\phi'_B)=\frac{2\sin^2\eta'}{\sin^2\theta'_B}-1,
\end{equation}

For SR3 model, we follow the treatment in Lan et al. (2019). The 3 dimensional isotropic random magnetic field of SR3 model can be described as $\hat{B}'=\hat{B}'_3=\sin\theta_r\cos\phi_r\hat{x}+\sin\theta_r\sin\phi_r\hat{y}+\cos\theta_r\hat{\beta}$. $\theta_r$ and $\phi_r$ are the polar and azimuthal angles of the random magnetic field in coordinate system $\hat{x}\hat{y}\hat{\beta}$. Using Eq. (7) of Lan et al. (2019), we will obtain the expressions for $\sin\theta'_B$, $\sin\phi'_B$ and $\cos\phi'_B$ for SR3 model. Then it can be proved that $\langle U'_p\rangle=0$, while it is hard to prove that $\langle Q'_p\rangle$ is also zero through parity of the integrand for SR3 model.

Local PAs for both SR2 and SR3 models depend on the sign of the $\langle Q'_p\rangle$, when $\langle Q'_p\rangle>0$, $\chi_p=\phi+3\pi/2$, when $\langle Q'_p\rangle<0$, then $\chi_p=\phi$ (Lan et al. 2019). Finally, it can be proved that the Stokes parameter $U_\nu$ is zero for both SR2 and SR3 models. In the following, we can see from our numerical results that PD for SR3 model is indeed 0, independent of all parameters.

\subsection{Compton Drag (CD) Model}
It is known that Compton scattering can induce polarization. In CD model, soft photons around the GRB jet will be up-scattered by the electrons in jet because of their relativistic bulk motion (Shaviv \& Dar 1995; Eichler \& Levinson 2003; Levinson \& Eichler 2004; Lazzati et al. 2004). Same as that in Toma et al. (2009), a nonthermal spectrum for the seed photons is assumed and also the seed photon field are assumed to be unpolarized and isotropic. Then it reads $A_0=(1+\cos^2\theta')/2$, $\Pi_p=(1-\cos^2\theta')/(1+\cos^2\theta')$ and $\chi_p=\phi+3\pi/2$, where $\cos\theta'=(\cos\theta-\beta)/(1-\beta\cos\theta)$. And also for CD model, its Stokes parameter $U_\nu$ is 0.

\label{lastpage}

\end{document}